\documentclass[aps,preprint,showpacs,superscriptaddress,groupedaddress]
{elsarticle}
\usepackage{bm}       
\usepackage{amssymb}
\usepackage{amsmath}     

\usepackage{hyperref} 
\usepackage{psfrag}
\usepackage[cp1250]{inputenc}

\usepackage{graphicx}

\DeclareMathAlphabet{\mathpzc}{OT1}{pzc}{m}{it}

\begin{document}

\title{\boldmath Implications of extreme flatness in a general f(R) theory\\}
\author[cracow]{Micha{\l} Artymowski} 

\author[warsaw]{Zygmunt Lalak} 

\author[warsaw]{Marek Lewicki}

\address[warsaw]{Institute of Theoretical Physics, Faculty of Physics, University of Warsaw ul. Pasteura 5, 02-093 Warsaw, Poland}
\address[cracow]{Institute of Physics, Jagiellonian University, {\L}ojasiewicza 11, 30-348 Krak{\'o}w, Poland}

\begin{abstract}
We discuss a modified gravity theory defined by $f(R) = \sum_{n}^{l} \alpha_n M^{2(1-n)} R^n$. We consider both finite and infinite number of terms in the series while requiring that the Einstein frame potential of the theory has a flat area around any of its stationary points. We show that the requirement of maximally flat stationary point leads to the existence of the saddle point (local maximum) for even (odd) $l$. In both cases for $l\to\infty$ one obtains the Starobinsky model with small, exponentially suppressed corrections. Besides the GR minimum the Einstein frame potential has an anti de Sitter vacuum. However we argue that the GR vacuum is absolutely stable and AdS cannot be reached neither via classical evolution nor via quantum tunnelling. Our results show that a Starobinsky-like model is the only possible realisation of $f(R)$ theory with an extremely flat area in the Einstein frame potential.
\end{abstract}

\date{\today}

\maketitle

\tableofcontents

\section{Introduction}

Cosmic inflation \cite{Lyth:1998xn,Liddle:2000dt,Mazumdar:2010sa} is a well established, consistent with the data \cite{Ade:2013uln}, theory of the early universe which predicts cosmic acceleration and generation of seeds of the large scale structure of the present universe. The inflationary universe can be obtained by introduction of additional fields or by modification of general relativity (GR), which is the possibility explored in this paper. The first theory of inflation is the Starobinsky model \cite{Starobinsky:1980te,Barrow:1988xh}, which is an $f(R)$ theory \cite{DeFelice:2010aj} with $R + R^2/6M^2$ Lagrangian density. In such a model the acceleration of space-time is generated in the empty universe, i.e. by the gravitational interaction itself. This comes from the fact that the homogeneous and isotropic $R^2$ model gives an exact de Sitter solution.
\\*

The $f(R)$ theory is one of the simplest generalisations of general relativity (GR). It is based on Lagrangian density $S = \frac{1}{2}\int d^4 \sqrt{-g}f(R)$ and it can be expressed using the so-called auxiliary field, which means that the Ricci scalar is treated as a independent scalar degree of freedom. In such a case one defines $Q$ by $Q:=R$ and the Jordan frame (JF) action is equal to 
\begin{equation}
S_{\text{\tiny JF}} = \int d^4x\sqrt{-g} \, \left(F(Q)\frac{R}{2} - U(Q) \right) \, ,
\end{equation}
where $F = df/dQ$ and $U(Q) = (QF(Q)-f(Q))/2$. $U$ is the Jordan frame (JF) potential, which is related to energy density of the field, but its derivative is not an effective force in the EOM of Q. For $F=1$ one recovers GR, which is usually positioned at $Q=0$. Note that the variation of the JF action with respect to $Q$ gives $F'(Q)(Q-R) = 0$. Therefore one obtains a constraint on $Q$ and $R$, which is valid whenever $F'(Q) \neq 0$. If $F' = 0$ the constraint is satisfied for any relation between $Q$ and $R$, { and therefore we lose one-to-one correspondence between the scalar picture and the original $f(R)$ theory.} The Jordan frame auxiliary field can also be defined as $\varphi:=F(R)$. This convention requires an exact form of the function $R=R(\varphi)$, however obtaining such a form is not always possible.
\\*

The same model can be expressed in the Einstein frame (EF), with the metric tensor defined as $\tilde{g}_{\mu\nu} = F(Q) g_{\mu\nu}$. This is a purely classical transformation of coordinates and results obtained in one frame are { classically} perfectly equivalent to the ones from another frame. The EF action is equal to
\begin{equation}
S_{\text{\tiny EF}} = \int d^4 \tilde{x} \sqrt{-\tilde{g}} \, \left(\frac{1}{2}\tilde{R}+\frac{1}{2}(\partial_\mu \phi)^2 - V(\phi)\right) \, ,
\end{equation}
where $\tilde{R}$, $\phi:=\sqrt{3/2}\,\log F$ and $V:=(RF-f)/(2F^2)$ are the EF Ricci scalar, auxiliary field and potential respectively. The EF potential should have a minimum at the GR vacuum, which is positioned at $\phi = 0$. The auxiliary field $\phi$ may be used as an inflaton or a source of the dark energy, which makes the $f(R)$ theory a powerful theoretical tool to solve problems of classical cosmology. 
\\*

A separate issue related to $f(R)$ inflation concerns loop corrections to the f(R) function. Note that the $R^2$ term in the Starobinsky model was originally motivated by one-loop correction to GR, which in principle could be extended into series of higher order loop corrections. In order to obtain quasi de Sitter evolution of space-time one needs a wide range of energies for which the $R^2 M^{-2}$ term dominates the Lagrangian density. This would require all higher order corrections (such as $R^3$, $R^4$ etc.) \cite{Ferrara:2013kca,Artymowski:2015pna} to be suppressed by a mass scale much bigger than $M$.
One naturally expects all higher order correction to GR to appear at the same energy scale if one wants to avoid  fine-tuning of coefficients of all higher order terms.
Thus, the influence of higher order corrections on the Starobinsky model may spoil the flatness of the Einstein frame potential and prevent the early universe from inflating.
The saddle-point inflation generated by higher order corrections to Starobinsky model was already analysed in \cite{Artymowski:2015pna,Artymowski:2015ida,Artymowski:2015mva}. In this paper we extend this analysis as follows: we assume that the Einstein frame potential of the $f(R)$ theory with higher order terms up to $R^l$ has a flat area around any of its stationary points. We do not assume where such a point is, we simply require that the potential is as flat as possible around it, namely that first $l-2$ derivatives of $V$ are equal to zero at this point, which gives certain relations between $\alpha_n$ coefficients. We describe implications of such extreme flatness. We also investigate the issue of stability of the GR vacuum of the model. { We stress that our point of view is that whatever corrections have been computed there exists an effective classical action that can be studied and this is what we do in this paper. Our approach also covers the case that higher order corrections come from the Taylor expansion of some fundamental, unknown $f(R)$ theory.}
\\*

In what follows we use the convention $8\pi G = M_{p}^{-2} = 1$, where $M_{p} = 2.435\times 10^{18}GeV$ is the reduced Planck mass.
\\*

The outline of the paper is as follows. In Sec. \ref{sec:stationary} we discuss the general form of $f(R)$ with a stationary point around which the Einstein frame potential is as flat as possible. In Sec. \ref{sec:stable} we investigate the stability of the GR vacuum and a possibility of quantum tunnelling to anti de Sitter vacuum. Finally we summarise in Sec. \ref{sec:concl}


\section{General $f(R)$ function with stationary point in Einstein frame} \label{sec:stationary}

\subsection{A stationary point with $k$ vanishing derivatives} \label{sec:kterms}

\begin{figure}[h!]

\includegraphics[height=5cm]{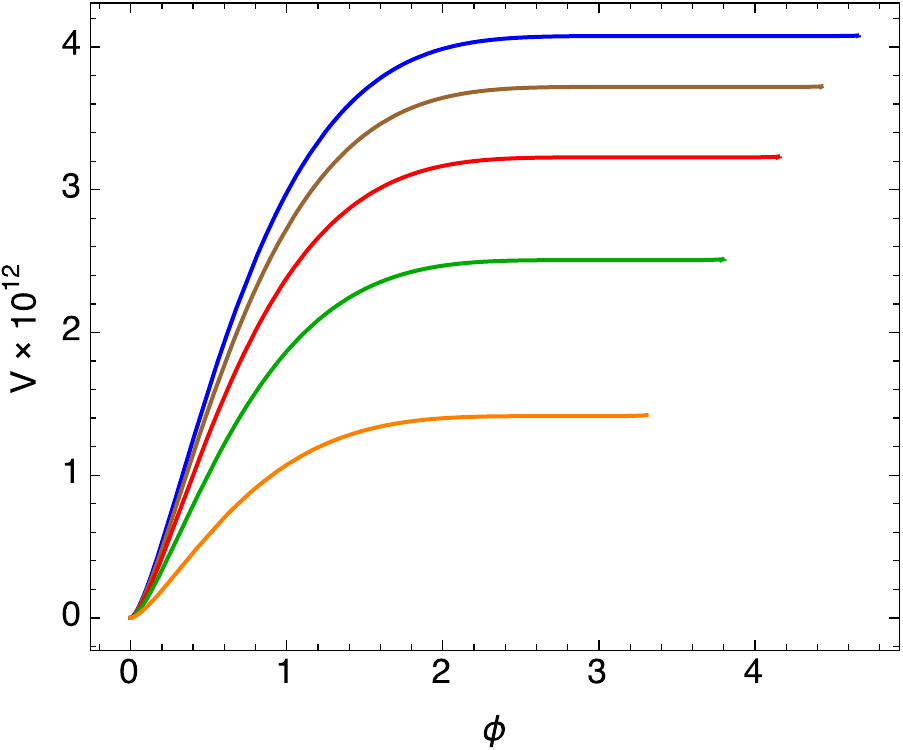}  
\includegraphics[height=5cm]{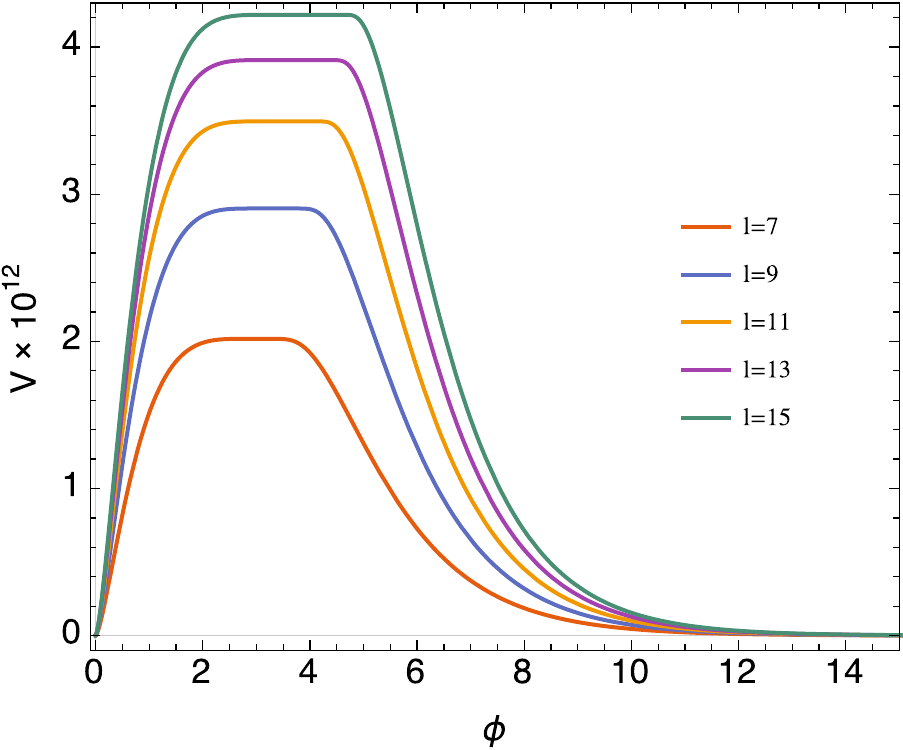}
\caption{\it Left panel: EF potential for the model (\ref{eq:fSum}) for $l=6$, $l=8$, $l=10$, $l=12$ and $l=14$ (orange, green, red, brown and blue lines respectively). The saddle point lies close to the right edge of the potential, beyond which one obtains a second branch of $V$, which leads to repulsive gravity. Right panel: plateaus around local maxima for several odd $l$. All potentials have runaway vacuum for big $\phi$. For both odd and even $l$ the width of the plateau grows with $l$, which leads to the infinite plateau for $l \to \infty$.}
\label{fig:M(l)}
\end{figure}

\begin{figure}[h!]
\includegraphics[height=5cm]{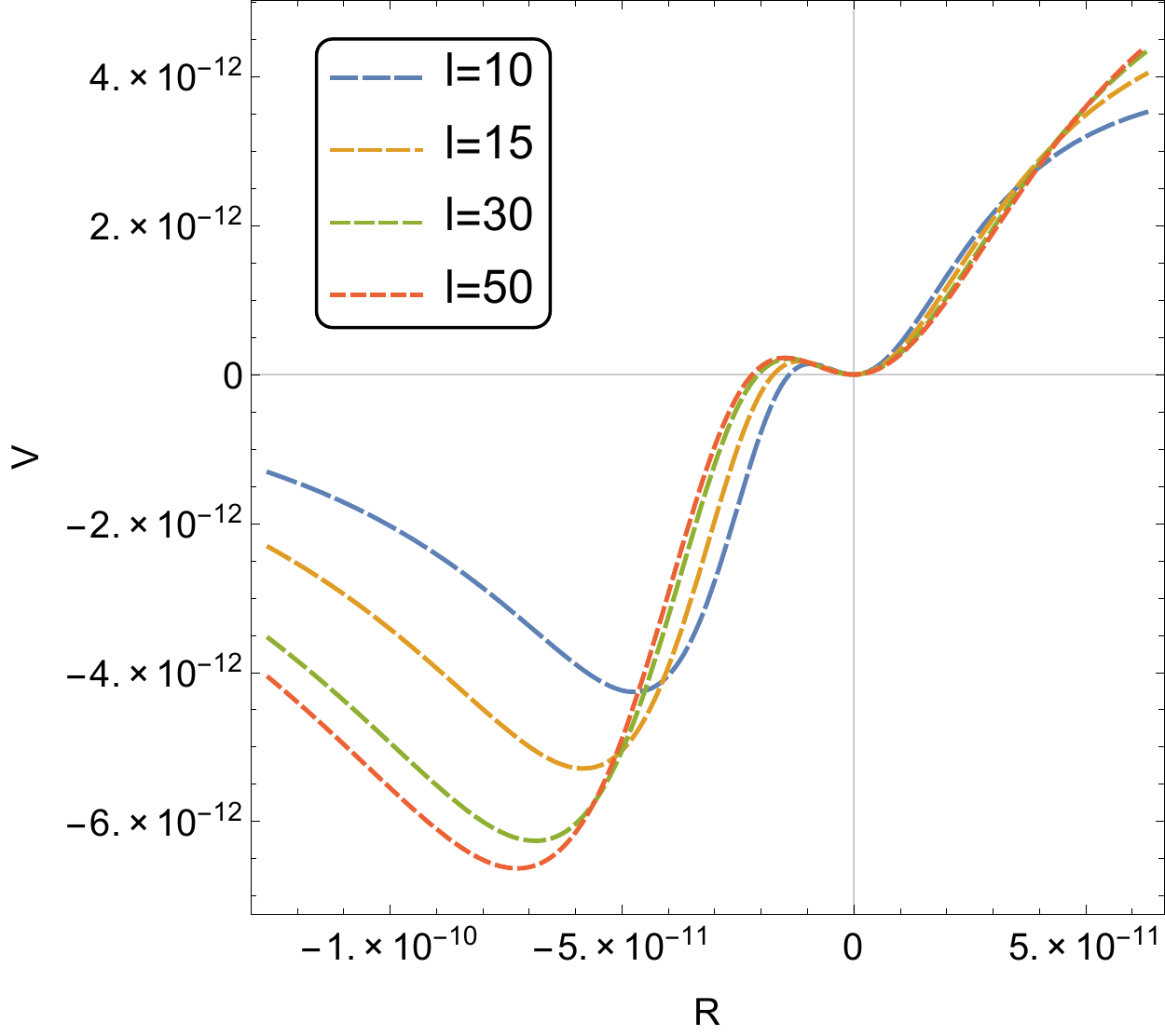} 
\includegraphics[height=5cm]{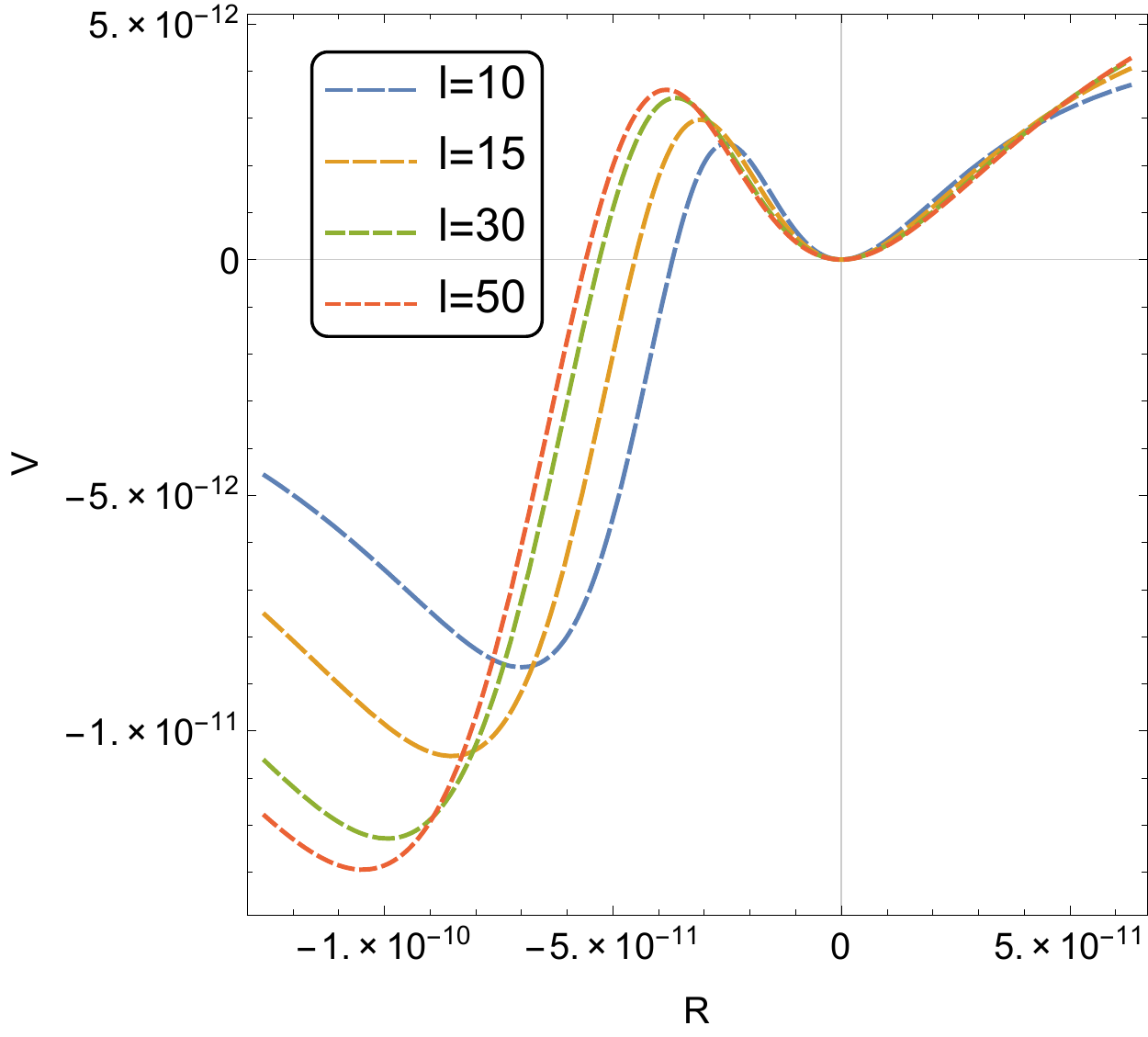}  
\caption{\it Einstein frame potential for different values of $l$ and $\alpha=1/2$ (left panel) and $\alpha=1$ (right panel).}
\label{fig:V(l)}
\end{figure}

As mentioned in the introduction we require the existence of a stationary point (i.e. extremum or saddle point) anywhere in the Einstein frame potential besides the minimum in $\phi = 0$, which is the GR vacuum. We require the potential around a stationary point to be as flat as possible for a given $f(R)$ function. We want to find out whether this requirement will determine the shape of $V(\phi)$ outside of the domain of a stationary point. Let us assume that the $f(R)$ function is the following sum of $R^n$ terms
\begin{equation}
f(R) =\sum _{n=1}^{l}\alpha_n\frac{R^n}{M^{2(n-1)}} \, ,\label{eq:sumF}
\end{equation}
where $l > 4$ is a natural number. In order to obtain the correct GR limit one requires $\alpha_1=1$.  Without any loss of generality one can choose $\alpha_3$ to be any positive constant, so for simplicity we set $\alpha_3=1$. Conditions $V_\phi = 0$ and $V_{\phi\phi} = 0$ are equivalent to $R F = 2f$ and $R F' = F$ for some $R = R_s$, where $R_s$ is a stationary point of the Einstein frame potential. All $\frac{d^k V}{d\phi^k}=0$ for $k>2$ are equivalent to $\frac{d^k f}{dR^k} = 0$. In case of a saddle point there is a deeper motivation to consider $R_s$ with many vanishing derivatives. The saddle point with $V_\phi = V_{\phi\phi} = 0$ and $V_{\phi\phi\phi} \neq 0$ gives spectral index $n_s \simeq 0.92$, which is inconsistent with the PLANCK data. On the other hand the saddle point with first $k$ derivatives vanishing, which was analysed in Ref. \cite{Hamada:2015wea},gives $1-n_s \simeq \frac{2k}{N_{\star}(k-1)}$, so one can fit the PLANCK data for $k$ of order of at least a few, if the pivot scale leaves the horizon when $\phi$ is close to the saddle point. For any given $l>2$ one can obtain maximally $l-2$ vanishing derivatives of the Einstein frame potential for some $\phi_s$. This comes from the fact that $R$ and $R^2$ automaticly satisfy $V^{(n)} = 0$ for any non-zero $n$. Assuming the maximal number of vanishing derivatives one obtains
\begin{equation}
R = R_s = \sqrt{p}\, M^2 \, ,\quad \text{where}\quad p= (l-1) \left(\frac{l}{2}-1\right)\, ,\label{eq:bfRs}
\end{equation}
where $R_s$ is a saddle point (local maximum) of $V$ for even (odd) $l$ respectively. The $\alpha_n$ coefficients satisfy
\begin{equation}
\alpha_n = (-1)^{n-1}\frac{2(l-3)!}{(l-n)!(n-1)!}p^{\frac{3-n}{2}}   \quad \text{for} \quad n = \{3,\ldots,l\} \, . \label{eq:alphan}
\end{equation}
Note that Eq. (\ref{eq:bfRs}) and (\ref{eq:alphan}) are completely independent of $\alpha_2$. For the odd (even) $l$ one finds $\alpha_l > 0$ ($\alpha_l < 0$) respectively. Thus for even $l$ one obtains $F<0$ for sufficiently big $R$ and the gravity becomes repulsive. { Note that $F = 0$ does not only separates the attractive and repulsive limit of gravity. It is also a pressure singularity, which cannot be passed by any trajectory in phase space.} Usually the maximal allowed $R$ is an order of magnitude bigger than $R_s$. For odd $l$ the potential is well defined for all $R>0$. The $\alpha_2$ is the only free parameter of the theory, since none of the conditions for the stationary point does not constrain it. Using Eq. (\ref{eq:sumF}) and (\ref{eq:alphan}) one obtains
\begin{equation}
f(R) = R+\frac{\alpha_2}{M^2} R^2+R\frac{\left(l M^2 \sqrt{p} R+M^4 p \left(\left(1-\frac{R}{M^2 \sqrt{p}}\right)^l-1\right)-(l-1) R^2\right)}{M^4 p-M^2 \sqrt{p} R}\, . \label{eq:fSum}
\end{equation} 
The Einstein frame potential for such an $f(R)$ theory has several troublesome points. The first one is the repulsive gravity limit for sufficiently big $R$ and even $l$. The second one is the possible instability of the GR vacuum at $R=0$. For $\alpha_2 = 0$ the only minimum is the anti de Sitter minimum at certain $R  < 0$. In order to create a potential barrier between the possibly unstable GR vacuum and anti de Sitter vacuum one needs $\alpha_2>0$, which will be analysed in detail in the section \ref{sec:stable}. We plot the Einstein frame potential as a function of $\phi$ and $R$ in Fig. \ref{fig:M(l)} and Fig. \ref{fig:V(l)} respectively.
\\*

In the Ref \cite{Artymowski:2015pna} we showed what are the features of the power spectrum of primordial inhomogeneities for the saddle-point case (even $l$). For odd $l$ the results are exactly the same, i.e. $r$ and $n_s$ have the same $l$ dependence as in the even $l$ scenario. For any finite $l$ there is an issue of initial conditions for inflation. For the saddle point inflation the $R$ cannot be too high in order to i) stay on the inflationary branch of the potential, ii) avoid the repulsive gravity regime. For odd $l$ the question is how has the field appeared on the plateau and why initial $\phi$ was smaller than $\phi_s$ (the opposite case would mean that the field rolls down towards the runaway vacuum). Note that for odd $l$ one can also obtain the topological inflation.

\subsection{Extensions to other scalar-tensor theories}

One could try to generalize this analysis into the Brans-Dicke theory. In Ref \cite{Artymowski:2015ida,Artymowski:2015mva} we have investigated the issue of higher order corrections to the JF potential in Brans-Dicke theory. We have proven that in the presence of higher order corrections one can still obtain flat areas of the Einstein frame potential, for instance around a saddle point. Nevertheless such a saddle point is not maximally flat, i.e. we required in Ref. \cite{Artymowski:2015ida,Artymowski:2015mva} only the first two derivatives to be zero, even though the number of free parameters enabled us to make the first three derivatives of $V$ vanish. This issue can be discussed for a general series of higher order corrections to the JF potential. It is easy to show that for the Jordan frame potential
\begin{equation}
U = \sum_{n=1} ^{l} \alpha_n (\varphi-1)^n 
\end{equation}
one cannot obtain a real, maximally flat stationary point in the Einstein frame. Therefore the idea presented in this paper does not work in the Brans-Dicke theory.
\\*

The other way to extend this analysis is to include negative powers of $R$ in the $f(R)$ power series. In particular such terms could be used as source of dark energy \cite{Amendola:2006we}. Nevertheless, negative powers of $R$ have several disadvantages. In Ref. \cite{Amendola:2006we} it was proven that For the $R - \alpha R^n$ models with $n<0$ one does not obtain the true dust domination era, which is inconsistent with astronomical observations. The other issue is the influence of negative powers of $R$ on the stability of GR vacuum - after including $n < 0$ in Eq. (\ref{eq:sumF}) the GR vacuum does not appear even for significant contribution of $\alpha_2$. Therefore we restrict our analysis to positive $n$.
\\*

Another form of scalar-tensor theory used to obtain inflationary potentials is the so-called induced inflation \cite{Kallosh:2014laa,Giudice:2014toa} with the following action 
\begin{equation}
S = \int d^4 \sqrt{-g} \left[\frac{1}{2}f(\varphi)R + \frac{1}{2}(\partial\varphi)^2-M^2 (f-1)^2 \right] \, .
\end{equation}
The EF potential takes the form $V = (1-1/f)2$ and therefore for big values of $f$ it has a Starobinsky-like plateau. Note, that for $f = 1 + \xi \varphi^2$ one recovers the Higgs inflation. This simple model was generalised into $f = 1+ \xi \varphi ^n$, which gives the same results as Higgs or Starobinsky inflation in the strong coupling limit \cite{Kallosh:2013tua}. For the following form of $f(\varphi)$ 
\begin{equation}
f(\varphi) = \xi \sum_{k=0}^n \, \lambda_k \, \varphi^k \, ,\label{eq:lambdak}
\end{equation}
the requirement of the existence of the maximally flat area around a stationary point of the EF potential gives
\begin{equation}
f(\varphi) = \lambda_0 + \frac{\xi}{n}\left(n \, \lambda_n\right)^{\frac{-1}{n-1}}\left(1 + \left(\left(n \, \lambda_n\right)^{\frac{1}{n-1}}\varphi - 1 \right)^n\right) \, .\label{eq:fgeneral}
\end{equation}
This form of $f(\varphi)$ contains all possible positive powers of $\varphi$ and besides the Starobinsky-like plateau for big $\varphi$ it predicts the existence of an additional plateau around the saddle point at $\varphi_s = \left(n \, \lambda_n\right)^{\frac{-1}{n-1}}$. Depending on values of $\lambda_0$ and $\lambda_n$ the two plateaus can be separated by the GR minimum or there can be a cascade of plateaus. Therefore this model can generate a multi-phase inflation, where each phase occurs at a different energy scale. { The same approach could be also used in the context of a scalar theory with minimal coupling to gravity. The results of this analysis} will be presented in our further work.

\subsection{The $l \to \infty$ limit} \label{sec:infterms}

Numerical analysis shows that in order to obtain correct normalisation of primordial inhomogeneities one needs $M=M(l,\alpha_2)$. Nevertheless for $l\to \infty$ one obtains $M \to M_o = M_o(\alpha_2)$ (where $M_o\sim 10^{-5}$ for $\alpha_2=0$), which implies $R_s \to \infty$ for $l \to \infty$. Hence for $l \gg 1$ one cannot obtain inflation close to stationary point. For $l\to\infty$ one obtains 
\begin{equation}
f(R) = R \left(e^{-\frac{\sqrt{2} R}{M_o^2}}+\frac{\sqrt{2}+\alpha_2}{M_o^2}R\right) \, . \label{eq:fSumAll}
\end{equation}
The result is a slightly modified Starobinsky model, which is very interesting, since we require the existence of a point around which the potential is perfectly flat. Note that one cannot obtain the analytical relation $\varphi = \varphi(R)$ for all $\alpha_2$. The GR vacuum of the Einstein frame potential of (\ref{eq:fSumAll}) does not exist without the explicit contribution of the $\alpha_2$ term. This comes from the fact that around $\phi = 0$ one finds 
\begin{equation}
V \simeq \frac{M^2}{12\alpha_2}\phi^2 - \frac{\left(2 \alpha_2^2+1\right) M^2}{12 \sqrt{6} \alpha_2^3}\phi^3 + \mathcal{O}(\phi^4).
\end{equation}
Without the $\alpha_2$ term all derivatives of $V$ around the GR vacuum are singular. Thus the positive $\alpha_2$ is needed to stabilise the GR vacuum at $\phi=0$. For $\alpha_2 < 0$ one finds $F<0$ for $R \lesssim \alpha_2 M_o^2$, so the GR limit lies in the range of repulsive gravity.
\\*

\begin{figure}[h!]
\includegraphics[height=4.7cm]{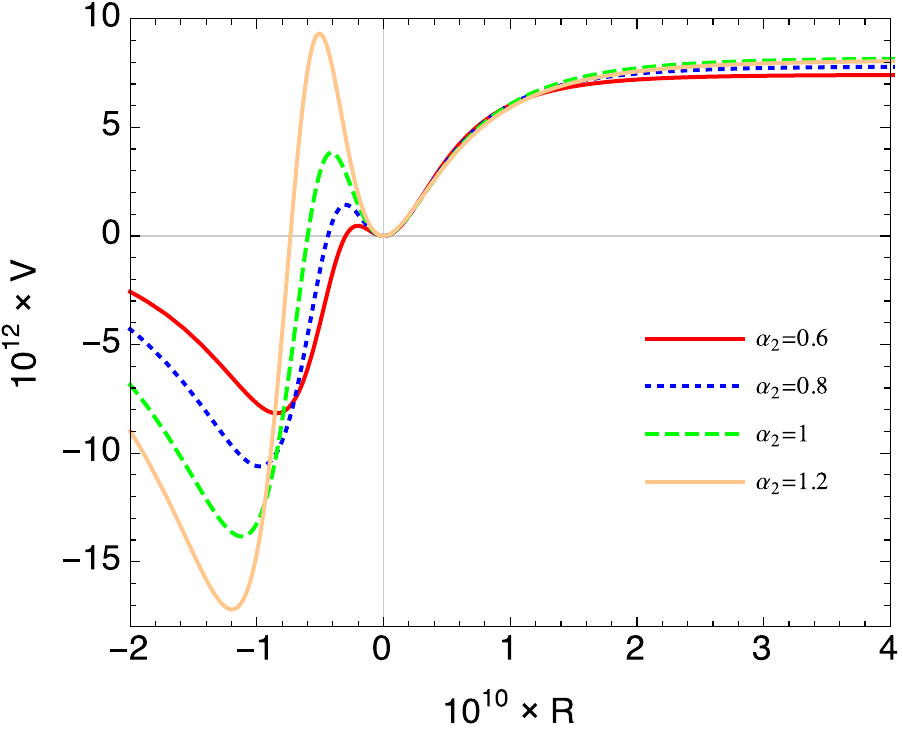} 
\includegraphics[height=4.7cm]{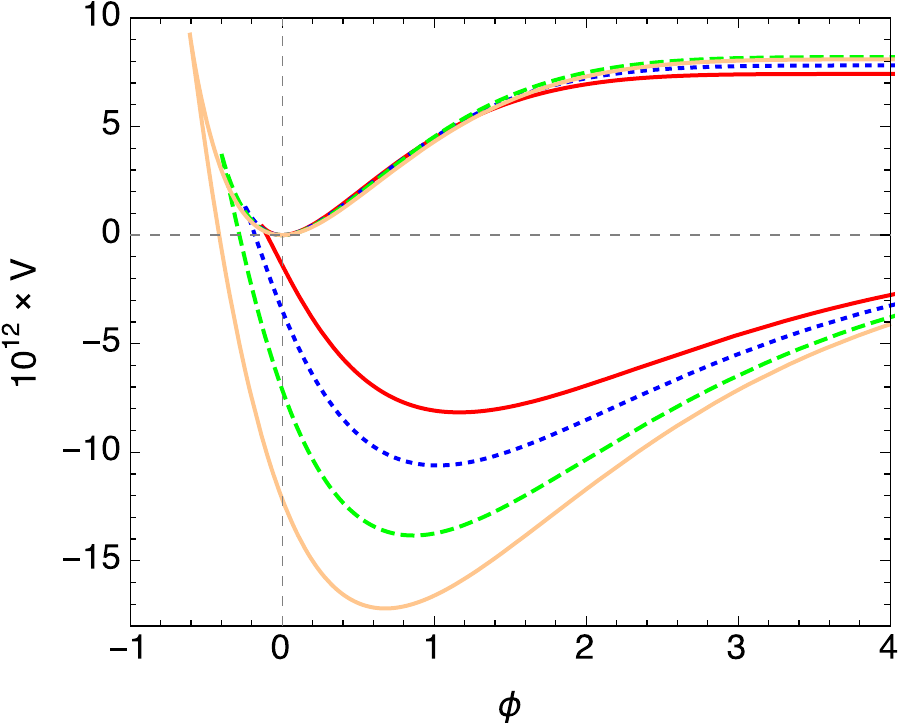}  
\caption{\it Left Panel: The Einstein frame potential as a function of the Ricci scalar. The GR minimum at $R=0$ seems to be unstable, due to the existence of the anti de Sitter vacuum. Right Panel: The Einstein frame potential $V$ as a function of the Einstein frame field $\phi$ for the model (\ref{eq:fSumAll}). Two branches of potential correspond to two solutions of $\phi = \sqrt{3/2}\log F(R)$.
}
\label{fig:V(alpha)}
\end{figure}

The numerical results for the power spectra for $N_\star = 60$ are shown in Fig. 6 of Ref \cite{Artymowski:2015pna}. As expected, for $\alpha \gg 1$ values of $M/\sqrt{\alpha}$, $r$ and $n_s$ assume limiting values of the Starobinsky theory. As shown in Fig. \ref{fig:V(alpha)} the potentials have two branches, which split at some $\phi = \phi_m$, where $\phi_m$ is the minimal value of $\phi$. The splitting point corresponds to the value of the Ricci scalar, for which $F' = 0$ and therefore the auxiliary field loses its relation to $R$. For $\alpha_2 = 0$ one obtains two branches of the potential which grow out from $\phi = 0$. Both of them do exist only for $\phi>0$ with no GR minimum. While increasing the value of $\alpha_2$ the splitting of branches moves towards $\phi < 0$ and the inflationary branch develops a minimum at $\phi = 0$. The splitting does not exists if one expresses $V$ as a function of $R$ or $Q$. Then one obtains two vacua (at $\phi=R=0$ and at $R = -M^2/\sqrt{2}$) separated by the maximum at $R = - M^2 (W (e^2 (\sqrt {2}\alpha + 2 )) - 2)/\sqrt {2}$, where $W$ is the Lambert function. For $\alpha_2 \gtrsim 1.2$ this maximum becomes a global maximum. 

\subsection{The self reproduction of the universe}

One can describe the evolution of the universe by the set of the classical equations of motion when quantum fluctuations of fields and metric remain small. This issue is especially important in the slow-roll regime for potentials, which are very flat or may assume very big values. Let us therefore investigate the slow-roll limit of the quantum and classical evolution of the scalaron (denoted as $\delta \phi$ and $\Delta\phi$ respectively) during one Hubble time $t_{\text{\tiny H}}$, which is the typical time scale for the inflationary universe. Then one obtains
\begin{equation}
\delta\phi < \Delta\phi \quad \Leftrightarrow \quad \left|(R F - f)^{3/2}\right| < 4\pi \left| F(2f - R F) \right| \, . \label{eq:cq}
\end{equation}
When the condition (\ref{eq:cq}) is not satisfied one obtains domination of quantum fluctuations over the classical evolution of $\phi$, so the field does not need to evolve towards its minimum. For $|\delta \phi| \gg |\Delta \phi|$ in a half of horizons generated during one Hubble time the value of $\phi$ would even grow! For the potentials with inflationary plateau this effect appears when the field is on the plateau, far enough from the GR minimum. The flatness of the plateau provides small $\Delta \phi$, which decreases while increasing $\phi$. Meanwhile $\delta \phi \propto \sqrt{V}$ remains almost $\phi$ independent, so at some $\phi > \phi_{q}$ one obtains $\delta\phi > \Delta\phi$.   
\\*

For the $R + \alpha_2 R^2/M^2$ Lagrangian density one finds $R_q \sim 8\pi M/\sqrt{\alpha_2}$ and therefore $\phi_q \sim \sqrt{\frac{3}{2}} \log \left(\frac{16 \pi  \sqrt{\alpha_2 }}{M}\right)$, which should be the $\alpha_2 \to \infty$ limit of the theory with higher order terms. The value of $R_q$ for the model (\ref{eq:fSumAll}) is plotted in the right panel of the Fig. \ref{fig:phidot}. Note that for $\alpha_2 < 10^6$ the quantum corrections start to dominate for much lower values of $R$ than in the Starobinsky model. Nevertheless $R_q$ is always at least an order of magnitude bigger than $R_\star$, which is the value of the Ricci scalar at which the pivot scale is leaving the horizon. Therefore the last 60 e-folds of inflation always remain determined by the classical evolution of the field.
\\*

{  The issue of quantum self-reproduction of the universe is considered in vast majority of inflationary theories. In this section we are not trying to solve it but rather to show that the classical evolution of the inflaton is valid at the crucial (i.e. observed) stage of inflation. Therefore it is sufficient that only one horizon would detach from eternal inflation, which would provide the graceful exit. Nevertheless one could argue that since vast majority of horizons are still inflating it is a priori highly improbable to live in the horizon that ever stopped inflating. Again, this problem is typical for inflation and not just for our model.}


\section{Classical and quantum stability of the GR vacuum} \label{sec:stable}

For both finite and infinite $l$ one obtains the EF potential with two vacua: i) the GR vacuum at $R = 0$, ii) the true vacuum of the model, which is the anti de Sitter minimum of the EF potential at $R = - M^2/\sqrt{2}$ (for $l \to \infty$) or at some negative $R$ (for finite $l$). One could ask whether it is possible to reach the true vacuum by classical evolution of the auxiliary field or by quantum tunnelling. In the first case the field could in principle overshoot the GR minimum while rolling down from the plateau. This could happen for a small $\alpha$, which would make the barrier between minima of the EF potential too low. Using the Ricci scalar as a scalar degree of freedom to discuss the quantum tunnelling \cite{Artymowski:2015mva,Coleman:1980aw}, we obtain the equation of motion
\begin{equation}
\ddot{R}+3\frac{\dot{a}}{a}\dot{R}=\frac{1}{3F'}(2f-RF-F''\dot{R}^2) \, , \label{eq:EOMQ}
\end{equation}  
where $F'=\partial F/\partial R$ and $\dot{R}=dR/dt$. Let us focus on the evolution of $R$ around maximum of the EF potential, which separates GR and anti de Sitter minima. Let us denote $R$ at the maximum as $R_{\max}$.
For any $l$ in our potential $F'(R_{\max})=0$, which means that for $R = R_{\max}$ one requires $2f - RF - F'' \dot{R}2 = 0$, to obtain a non=singular solution. 
Let us assume that $l \gg 1$ and $\alpha < 1$, which is a generic case for the issue of instability of the GR vacuum.
Then $R_{\max} \sim -\alpha M^2 /3$ and $(2f-R F)/F'' (R = R_{\max}) \sim -\alpha M^2 / 18$. In fact for all $\alpha$ and $l$ around $R \simeq R_{\max}$ one finds $\dot{R}^2=(2f-RF)/F'' < 0$, and so, Eq. (\ref{eq:EOMQ}) does not have real non-singular solutions connecting the two vacua. Only real solutions are physical and therefore all potentials, which require complex solutions to allow $R$ to reach $R_{\max}$ are excluded. In particular we should limit our parameter space to those potentials, which do not allow us to reach $R_{\max}$ starting from the plateau with the slow-roll initial conditions for $R$. Taking into account the $\alpha_2$ dependence of $M$ we have calculated the minimal value of $\alpha_2$ (denoted as $\alpha_{\min}$), which prevents $R$ from reaching $R_{\max}$ and therefore overshooting the GR minimum. The result is plotted in Fig. 8 in Ref. \cite{Artymowski:2015pna}. 
\\*

Another way to reach anti de Sitter vacuum would be to satisfy $2f - RF - F'' \dot{R}^2 = 0$ with $\dot{R} = 0$ at $R = R_{\max}$. This condition can be satisfied for 
\begin{equation}\label{eqn:topalpha}
\alpha_2 \simeq \frac{1}{2\sqrt{p}} \left( \left(\frac{l-1}{l-2}\right)^{l-2} (3 l-4)-2 (l-1) \right)\, ,
\end{equation}
which in the $l \to \infty$ limit gives $\alpha = (3e-2)/\sqrt{2}$. In this case one could possibly obtain an enormously finely tuned solution in which the curvature freezes at $R = R_{\max}$. Some quantum fluctuation could then push it towards anti de Sitter vacuum. Nevertheless values of $\alpha$ which allow for such a solution are $\sim 5$ times bigger than $\alpha_{\min}$ mentioned already in this section. Therefore the maximum would be too high, and we would not be able to reach it assuming slow-roll initial conditions on the plateau. Initial conditions on the plateau beyond the slow-roll approximation would mean that inflation did not commence and there is no hope to reconcile such solutions with experimental data.
\\*

\begin{figure}[h!]
\begin{center}
\includegraphics[height=3.6cm]{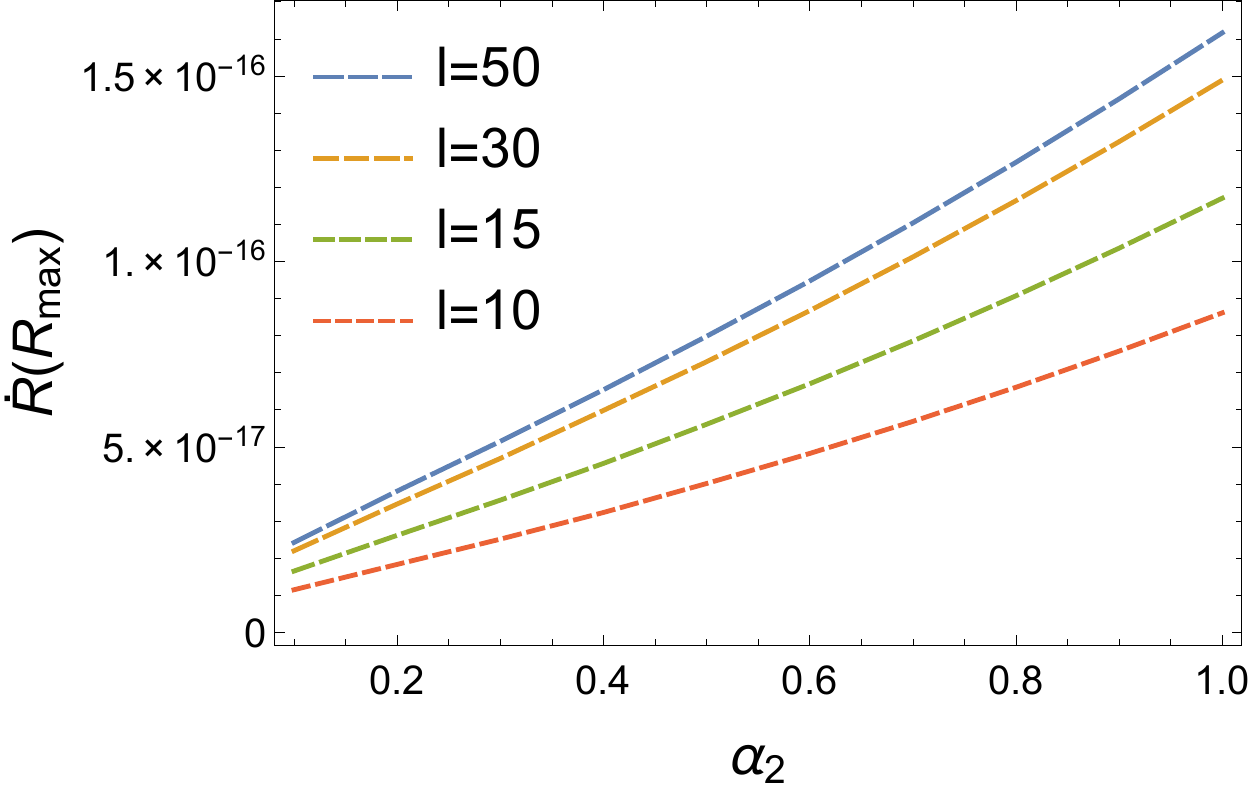} 
\hspace{0.4cm}
\includegraphics[height=3.6cm]{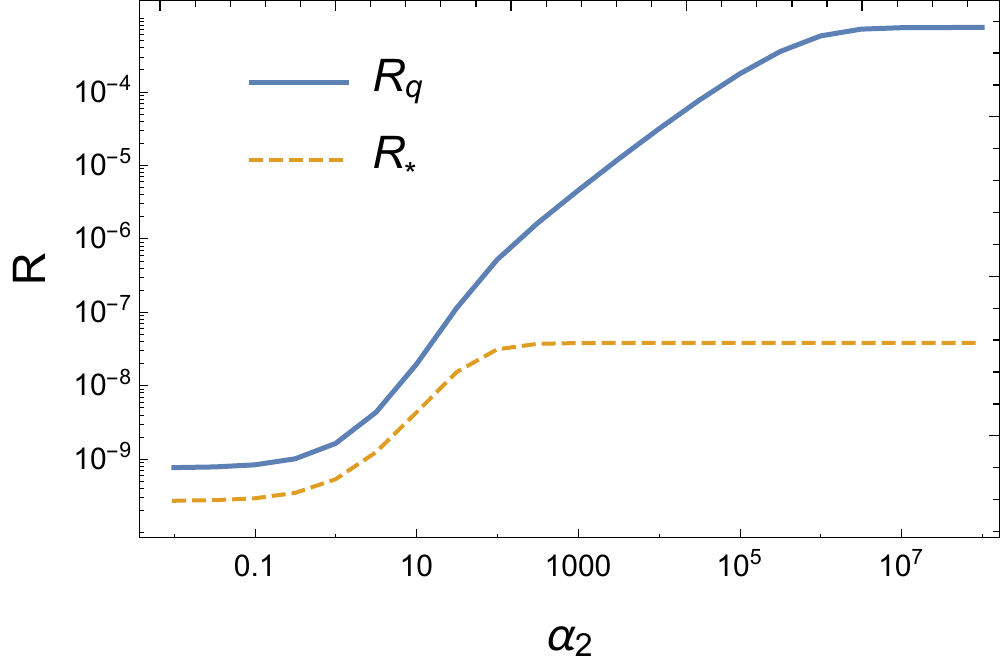} 
\end{center} 
\caption{\it Left panel: Values of $\dot{\varphi}$ at the maximum of the Einstein frame potential for which a non singular solution to the euclidean equation of motion exists. Values of $\dot{\varphi}$ for a physical time and the same $l$ and $\alpha$ are purely imaginary. right panel: $R_q$ and $R_*$ as a functions of $\alpha_2$ for $l \to \infty$ model. For $\alpha_2 \gtrsim 10^6$ one recovers the result of the Starobinsky theory.} 
\label{fig:phidot}
\end{figure}

In euclidean version of the theory the EOM, we need to solve while discussing quantum tunnelling, reads
 \begin{equation}
\ddot{R}+3\frac{\dot{a}}{a}\dot{R}=\frac{1}{3F'}(-2f+RF-F''\dot{R}^2) \, . \label{eq:EuclideanEOMQ}
\end{equation} 
Now with the sign of potential term from (\ref{eq:EOMQ}) changed, the RHS of EOM is no longer complex around $R = R_{\max}$. In fact there exists a single value of $\dot{R}_{max}$ for each $l$ and $\alpha$ which results in a nonsingular solution.
This value is shown is shown in Fig. \ref{fig:phidot}. 
Thus, it is possible to obtain a real solution of Eq. (\ref{eq:EuclideanEOMQ}), which passes $R_{\max}$, but it does not imply that quantum tunnelling of $R$ into the anti de Sitter vacuum is possible.
We also need to satisfy specific boundary conditions for the solution to represent an appearing bubble of the true vacuum. Most importantly we require that for $t\to \infty$ the solution asymptotes to our background, the GR vacuum. In order to check whether our single possible value of $\dot{R}_{max}$ represents a tunnelling solution, we solve the EOM from $R_{max}$ towards the GR vacuum and check whether the boundary condition is fulfilled. Generically this is not the case and even if a there exists a value of $\alpha$ for a given $l$ for which we obtain a solution this excludes only a single point in the parameter space. Thus we conclude that the GR vacuum is in general stable also with respect to quantum tunnelling.    

Vacuum stability in a model with a potential which splits in to two branches was recently discussed in \cite{Ibanez:2015fcv} with the application of the thin-wall approximation. This approximation does not require finding the solution to EOM and is more related to the other process which can render the vacuum  unstable, that is the Hawking-Moss transition \cite{Hawking:1981fz}. Essentially it is a temperature effect in which the system is excited to an unstable configuration on top of the barrier separating the two vacua. The role of temperature is played by Hawking-Gibbons temperature $T_{dS}=H/(2\pi)\approx\sqrt{V/3}/(2\pi)$ \cite{Brown:2007sd}.
Action of the HM instanton is the difference between the action of homogeneous solution of field in the false vacuum and on the top of  the barrier at $\phi_{\max}$
\begin{equation}\label{eq:tau}
S_{\textrm{HM}}=S_{\max}-S_{GR}
= 24  \pi^2 \left(\frac{1 }{ V_{\max}} - \frac{1}{V_{ GR}}\right).
\end{equation} 
To reach the top of the barrier we would need to satisfy \eqref{eqn:topalpha}.
However we can already see a problem since the decay probability is as usual exponentially suppressed by the action
\begin{equation}\label{eq:tau}
\Gamma \propto e^{-S_{HM}}.
\end{equation} 
In our model $ V_{\max}$ is of the order of $10^{-13}$ while $V_{GR}\approx 10^{-120}$ (both in Planck units), and the resulting action is enormous $S\approx 10^{122}$ leaving the decay practically impossible. 
Of course we could tune the model parameters by lowering $\alpha_2$ to lower the barrier to $\alpha_2 \ll 1$. However a more stringent constraint comes from requiring that the field does not reach the top of the barrier during classical evolution from normal slow-roll initial conditions needed for inflation. 

\section{Conclusions}\label{sec:concl}

In this paper we investigate the issue of very general higher order corrections to an $f(R)$ model in the context of the slow-roll inflation. In Sec. \ref{sec:stationary} we consider $f(R) = \sum_{n}^{l} \alpha_n M^{2(1-n)} R^n$ and we require the existence of a maximally flat area around a stationary point $\phi_s$ in the Einstein frame potential $V$. This requirement is equivalent to vanishing of the first $l-2$ derivatives of $V$ at $\phi = \phi_s$, which gives us values of all $\alpha_n$ coefficients for $n\geq 3$. The stationary point appears to be a saddle point (local maximum) for even (odd) $l$ respectively. In both cases power spectra of primordial inhomogeneities are consistent with PLANCK for $l\geq 10$. The $\alpha_2 R^2$ term is not constrained by the flatness of $V$ at $\phi = \phi_s$ and therefore $\alpha_2$ is a free parameter of a theory. A contribution of the $R^2$ term is needed in order to obtain GR vacuum of $V$.
\\*

In the $l \to \infty$ limit one obtains $f(R) = R (e^{-\sqrt{2} R/M_o^2}+(\sqrt{2}+\alpha_2)R/M_o^2)$, which is the Starobinsky model with an exponentially suppressed deviation. In such a model $V$ has two minima, namely the possibly unstable GR minimum for $\alpha_2>0$ and an anti de-Sitter minimum at $R = -M^ / \sqrt{2}$. Let us stress that again - we have started from the most general form of $f(R)$ with all possible $R^n$ terms (for $n>0$) and we have required that somewhere on $V$ there is a stationary point around which the potential is as flat as possible. Even though we have not assumed anything about other parts of potential we have obtained a Starobinsky-like model with flat inflationary plateau and with predictions consistent with the PLANCK data. 
\\*

The existence of the anti de Sitter vacuum rises a possibility of an instability of the GR minimum, which we analyse in Sec. \ref{sec:stable}. Minima are separated by a local maximum of the Einstein frame potential at $R = R_{\max}$. For $R = R_{\max}$ one finds $F' = 0$, which causes discontinuity of $V_\phi$. In order to satisfy Friedmann equations at $R = R_{\max}$ one needs complex values of curvature. This makes every solution, which reaches $R_{\max}$ unphysical. In order to avoid this one requires $\alpha_2 \gtrsim 0.7$, so the maximum is too high to be reach by the scalaron with slow-roll initial conditions on the plateau. For the Euclidean time the solution, which passes $R_{\max}$ can be real. Nevertheless, for every set of $\alpha$ and $l$ one obtains just two trajectories, which passes $R_{\max}$, so quantum tunnelling is very improbable. The GR vacuum of the model is therefore perfectly stable.

\section*{Acknowledgements}
This work was partially supported by the National Science Centre under research grants DEC-2012/04/A/ST2/00099 and DEC-2014/13/N/ST2/02712.
ML was supported by the Polish National Science Centre under doctoral scholarship number 2015/16/T/ST2/00527. MA was supported by National Science Centre grant FUGA UMO-2014/12/S/ST2/00243. 
ZL thanks DESY Theory Group for hospitality. This work was supported by the German Science Foundation (DFG) within the Collaborative Research Center (SFB) 676 Particles, Strings and the Early Universe.\\


\end{document}